**Nonlinear dynamics of autonomous vehicles with limits on acceleration**

L. C. Davis, 10244 Normandy Dr., Plymouth, MI 48170, United States

Abstract

The stability of autonomous vehicle platoons with limits on acceleration and deceleration is determined. If the leading-vehicle acceleration remains within the limits, all vehicles in the platoon remain within the limits when the relative-velocity feedback coefficient is equal to the headway time constant $[k = 1/h]$. Furthermore, if the sensitivity $\alpha > 1/h$, no collisions occur. String stability for small perturbations is assumed and the initial condition is taken as the equilibrium state. Other values of $k$ and $\alpha$ that give stability with no collisions are found from simulations. For vehicles with non-negligible mechanical response, simulations indicate that the acceleration-feedback-control gain might have to be dynamically adjusted to obtain optimal performance as the response time changes with engine speed. Stability is demonstrated for some perturbations that cause initial acceleration or deceleration greater than the limits, yet do not cause collisions.

1. **Introduction**

Autonomous vehicles or adaptive cruise control (ACC) vehicles could be a significant factor in future transportation systems. Various authors have shown that if ~30% of vehicles had ACC, the formation of jams in heavy traffic could be eliminated [1-10]. The ACC vehicles are assumed to be governed by control parameters that give string stability [11].

As the literature on string stability is large, no attempt to summarize all the papers will be given here. The reader is referred to several good references on traffic and ACC [12-16]. However, I will describe the salient developments as they relate to the present work. The first is the constant headway time policy that requires the control system to maintain the headway between two vehicles as $hv$, where $h$ is the headway time constant and $v$ is the velocity [17-19].

The control algorithm for a vehicle's acceleration is (assuming instantaneous mechanical response) $a = \frac{\alpha}{h}(\Delta x - D - hv) + k\Delta v$, where $\alpha$ is the sensitivity, $k$ is the gain for relative velocity feedback control, $D$ is the length of the vehicle plus a safety margin, $\Delta x$ is the center-to-center distance between vehicles, and $\Delta v$ is relative velocity $\frac{d\Delta x}{dt}$ [20]. Liang and Peng showed that string stability is attained if $\alpha + 2k > \frac{2}{h}$ [11].

If the mechanical response of the vehicle is characterized by a first-order time constant $\tau$, the maximum allowed $\tau$ occurs when $k = \frac{1}{h}$ independent of $\alpha$ and is given by $\tau_{max} = \frac{h}{2}$ [20]. String stability has been also analyzed for an explicit delay time $t_d$ by Orosz, Moehlis and Bullo [21] as well as for more general mechanical responses [22, 23]. To date, however, the effects of comfortable limits on acceleration and deceleration have not been examined in depth. The purpose of this paper is to examine platoon stability with such limits.

The only comparable model that has maximum acceleration and deceleration is the Intelligent Driver Model (IDM), which can be used as a type of ACC model [24]. The effects of mechanical response are not included in the IDM and the maximum deceleration possible implies that brakes are activated. The time delay for effective brake activation, which can be an important factor, is not included in the model or in the present analysis.

The paper is organized as follows. Sec. 2 is about a simple model for which an analytic result is proven. Sec. 3 contains simulations illustrating the results of Sec. 2. Sec. 4 is devoted to avoiding collisions that the limits might cause. Sec. 5 reports simulations for the more general model where mechanical response is included. Sec. 6 pertains to non-equilibrium initial states resulting in acceleration beyond the limits, yet cause no collisions. Conclusions are drawn in Sec. 7.

## 2. Simple Model

In this section I consider a simple model that illustrates the consequences of imposing limits on the acceleration on an otherwise string-stable platoon. For simplicity, I take $\tau = t_d = 0$ (instantaneous mechanical response) and let the maximum acceleration and deceleration be $a_{max}$. The maximum deceleration comes from the powertrain when the power is reduced rather than from braking. (Otherwise a braking algorithm treating the dynamics of activation must be included in the model.) All vehicles are taken to be identical and are labeled by $n = 1,2...$ with $n = 1$ being the first vehicle of the platoon. (I take $n = 0$ to be the leading vehicle whose velocity profile is to be specified.) The equation of motion for vehicle $n$ is therefore

$$a_n(t) = A_n(t), \qquad |A_n(t)| < a_{max}, \qquad (2.1a)$$

$$= a_{max} \, sgn[A_n(t)], \quad otherwise, \qquad (2.1b)$$

with

$$A_n(t) = \alpha\big[V_{op}(x_{n-1}(t) - x_n(t) - D) - v_n(t)\big] + k[v_{n-1}(t) - v_n(t)], \qquad (2.2)$$

where

$$V_{op}(z) = v_{max}, \qquad z \geq hv_{max}, \qquad (2.3a)$$

$$= \frac{z}{h}, \qquad 0 < z < hv_{max}, \qquad (2.3b)$$

$$= 0, \qquad z \leq 0. \qquad (2.3c)$$

The unconstrained acceleration is $A_n(t)$ and $V_{op}(z)$ is the optimal velocity which is limited by a maximum velocity (e.g., the speed limit) $v_{max}$.

Next I calculate the acceleration of vehicle $n$ when vehicle $n$ -1 does not exceed the limits on acceleration. For simplicity I take $k = \frac{1}{h}$. At time $t$ =0 the system is in equilibrium. So for $n$ = 1,2…

$$v_n(0) = v_{n-1}(0), \qquad (2.4a)$$

$$x_{n-1}(0) - x_n(0) - D = hv_n(0). \qquad (2.4b)$$

Assume that

$$-a_{max} \leq a_{n-1}(t) \leq a_{max}. \qquad (2.5)$$

Because the system is initially in equilibrium and $a_n(0) = 0$, the equation of motion is (assuming all velocities are positive, but less than $v_{max}$)

$$\ddot{x}_n + \left(\alpha + \frac{1}{h}\right)\dot{x}_n + \frac{\alpha}{h}x_n = \frac{\alpha}{h}\left(x_{n-1} - D + \frac{1}{\alpha}v_{n-1}\right). \tag{2.6}$$

Eq. (2.6) remains valid until $|a_n(t)|$ exceeds $a_{max}$ (if it ever does).

Let

$$y_n = \dot{x}_n + \alpha x_n, \tag{2.7}$$

so that

$$h\dot{y}_n + y_n = y_{n-1} - \alpha D. \tag{2.8}$$

The solution to Eq. (2.8) is

$$y_n(t) = -\alpha D + [y_n(0) + \alpha D]e^{-t/h} + \int_0^t e^{-\frac{t-t'}{h}} y_{n-1}(t') \frac{dt'}{h}, \tag{2.9a}$$

$$y_n(0) = v_n(0) + \alpha x_n(0). \tag{2.9b}$$

Now (rewriting Eq. (2.7))

$$\dot{x}_n + \alpha x_n = y_n(t). \tag{2.10}$$

The solution to Eq. (2.10) is

$$x_n(t) = x_n(0)e^{-\alpha t} + \int_0^t e^{-\alpha(t-t')} y_n(t')dt', \tag{2.11a}$$

$$= -D + [x_n(0) + D]e^{-\alpha t} + [y_n(0) + \alpha D]\left[\frac{e^{-\frac{t}{h}} - e^{-\alpha t}}{\alpha - \frac{1}{h}}\right] + w(t), \tag{2.11b}$$

where

$$w(t) = \int_0^t e^{-\alpha(t-t')} dt' \int_0^{t'} e^{-\frac{t'-t''}{h}} y_{n-1}(t'') \frac{dt''}{h}, \tag{2.12a}$$

$$= \int_0^t e^{-\alpha(t-t')} dt' \int_0^{t'} e^{-\frac{t'-t''}{h}} [\dot{x}_{n-1}(t'') + \alpha x_{n-1}(t'')] \frac{dt''}{h}. \tag{2.12b}$$

A change in the order of integration gives

$$w(t) = \int_0^t [\dot{x}_{n-1}(t') + \alpha x_{n-1}(t')] \frac{1}{h\left(\alpha - \frac{1}{h}\right)} \left[e^{-\frac{t-t'}{h}} - e^{-\alpha(t-t')}\right] dt'. \quad (2.13)$$

The part of Eq. (2.13) involving $x_{n-1}(t')$ can be integrated by parts to give two terms:

$$w_1(t) = \frac{\alpha}{\alpha - \frac{1}{h}} \left\{ x_{n-1}(t) - x_{n-1}(0)e^{-\frac{t}{h}} - \int_0^t \dot{x}_{n-1}(t') e^{-\frac{t-t'}{h}} dt' \right\} \quad (2.14)$$

and

$$w_2(t) = \frac{-1}{h\left(\alpha - \frac{1}{h}\right)} \left\{ x_{n-1}(t) - x_{n-1}(0)e^{-\alpha t} - \int_0^t \dot{x}_{n-1}(t') e^{-\alpha(t-t')} dt' \right\}. \quad (2.15)$$

After combining Eqs. (2.14) and Eqs. (2.15), I have

$$w(t) = x_{n-1}(t) - x_{n-1}(0) \left[ \frac{\alpha e^{-\frac{t}{h}} - \frac{1}{h} e^{-\alpha t}}{\left(\alpha - \frac{1}{h}\right)} \right] - \int_0^t \dot{x}_{n-1}(t') e^{-\frac{t-t'}{h}} dt' \quad (2.16)$$

and

$$x_n(t) = x_{n-1}(t) - D + \frac{v_n(0) - \alpha H_n(0)}{\left(\alpha - \frac{1}{h}\right)} e^{-\frac{t}{h}} + \frac{H_n(0) - h v_n(0)}{h\left(\alpha - \frac{1}{h}\right)} e^{-\alpha t} - \int_0^t \dot{x}_{n-1}(t') e^{-\frac{t-t'}{h}} dt', \quad (2.17)$$

where

$$H_n(0) = x_{n-1}(0) - x_n(0) - D. \quad (2.18)$$

Because the initial condition assumed is

$$H_n(0) = h v_n(0), \quad (2.19)$$

Eq. (2.17) simplifies to

$$x_n(t) = x_{n-1}(t) - D - h v_n(0) e^{-\frac{t}{h}} - \int_0^t v_{n-1}(t') e^{-\frac{t-t'}{h}} dt'. \quad (2.20)$$

Taking a time derivative of Eq. (2.20) gives the velocity

$$v_n(t) = v_n(0) e^{-\frac{t}{h}} + \frac{1}{h} \int_0^t v_{n-1}(t') e^{-\frac{t-t'}{h}} dt'. \quad (2.21)$$

A further time derivative gives the acceleration

$$a_n(t) = \frac{1}{h}\int_0^t a_{n-1}(t')e^{-\frac{t-t'}{h}}dt'. \tag{2.22}$$

Now

$$|a_n(t)| = \left|\frac{1}{h}\int_0^t a_{n-1}(t')e^{-\frac{t-t'}{h}}dt'\right|, \tag{2.23a}$$

$$\leq \frac{1}{h}\int_0^t |a_{n-1}(t')|e^{-\frac{t-t'}{h}}dt', \tag{2.23b}$$

$$\leq a_{max}. \tag{2.23c}$$

Hence, if the platoon is initially at equilibrium and the acceleration of the leading vehicle remains within the limits, the magnitude of the acceleration of each vehicle in the rest of the platoon will not exceed $a_{max}$. The string stability condition therefore applies because the limits on acceleration are not reached and consequently this condition ensures that the platoon is stable, unless collisions occur.

3. Simulations

For the platoon to be stable, vehicles must not collide with the vehicle immediately in front, *i.e.,*

$$x_{n-1}(t) - x_n(t) > D, \tag{3.1}$$

for all time. In this section I use simulations to determine the conditions for no collisions.

The acceleration $a_l$ of the leading vehicle (which is followed by the platoon) is chosen to be periodic with alternating values of $\pm a_{max}$. During the first period, which is of duration $T_p$,

$$a_l = a_{max}, \ 0 \leq t < \frac{T_p}{2}, \tag{3.2a}$$

$$a_l = -a_{max}, \ \frac{T_p}{2} \leq t < T_p. \tag{3.2b}$$

The velocity and acceleration are shown in the Fig. 1 for $T_p = 20$ s and initial velocity $v_l(0) = 16$ m/s. Simulations were carried out with $h = 1$ s and $a_{max} = 1$ m/s² for a range of values of $\alpha$ and $k$. The calculations were done for 300 s and 10 vehicles. The time increment (update time) is 0.01 s. At $t = 0$, for all vehicles following the leading vehicle

$$v_n(0) = v_l(0) \tag{3.3a}$$

and

$$x_n(0) = x_{n-1}(0) - hv_l(0) - D. \tag{3.3b}$$

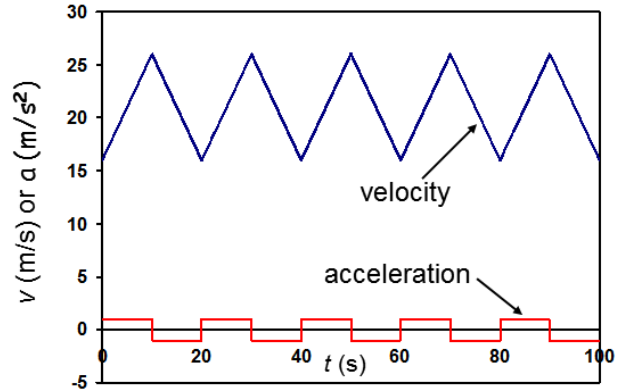

Fig. 1. The velocity (blue curve) and acceleration (red curve) of the vehicle leading a platoon. The vehicle alternately accelerates at $a_{max}$ = 1 m/s² and decelerates at $-a_{max}$. The period of the oscillations is $T_p$ = 20 s. The platoon vehicles labeled $n$ = 1, 2 … follow the leading vehicle.

Fig. 2 shows the results for a practical range of $\alpha, k$ values (not just $k = \frac{1}{h}$). The blue line is the string stability boundary. The red squares show the highest value of $k$ for each $\alpha$ at which a collision occurs. Above these $k$ values, no collisions occur and the acceleration and deceleration of the 10$^{th}$ (and larger $n$) vehicle are less than $a_{max}$. These simulations indicate the platoon is stable and the amplitude of acceleration decreases as $n$ increases.

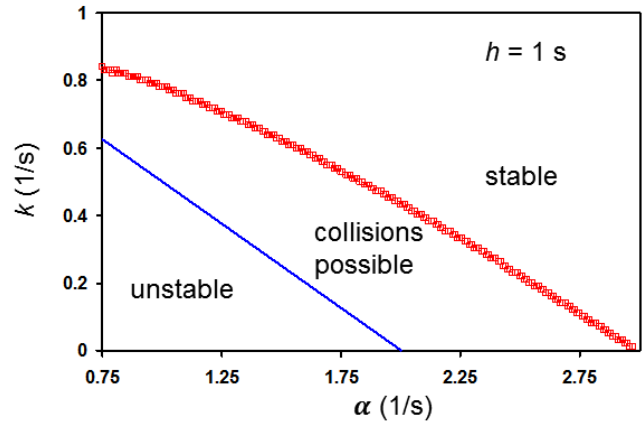

Fig. 2. The regions in $\alpha - k$ space where the motion of the platoon following the leading vehicle (Fig. 1) is stable, involves collisions (headway is zero or less), and is unstable. The blue line is the string stability condition $\alpha + 2k > \frac{2}{h}$. The red squares are simulation results for the first $k$ value (as $k$ is decreased) at each $\alpha$ for which a collision occurs. The initial condition is the equilibrium state. (All vehicles initially have velocity 16 m/s and headway 16 m.) The headway time constant $h$ = 1 s for all simulations. The vehicle mechanical response is taken to be instantaneous.

4. Avoiding Collisions

In this section I further examine the requirements for avoiding collisions. First, I find the platoon motion when the platoon follows a leading vehicle whose motion is given by

$$v_l(t) = v_0 - a_{max}t. \tag{4.1}$$

At $t = 0$

$$v_1(0) = v_0, \tag{4.2a}$$

$$x_l(0) = x_1(0) + hv_0 + D. \tag{4.2b}$$

The first vehicle in the platoon obeys the following equation (let $x_1(t) - x_1(0) = x$) until the deceleration reaches $a_{max}$:

$$\ddot{x} + (\alpha + k)\dot{x} + \frac{\alpha}{h}x = (\alpha + k)v_0 + \frac{\alpha}{h}\left[v_0 t - \frac{a_{max}}{2}t^2\right] - ka_{max}t. \tag{4.3}$$

The general solution to Eq. (4.3) is

$$x(t) = -\frac{a_{max}}{2}t^2 + ct + s + C_+ e^{\lambda_+ t} + C_- e^{\lambda_- t}, \tag{4.4}$$

where

$$c = v_0 + ha_{max}, \tag{4.5}$$

and

$$s = -\frac{h}{\alpha}a_{max}[h(\alpha + k) - 1]. \tag{4.6}$$

The homogenous solution has two roots of the characteristic equation

$$\lambda_\pm = \frac{1}{2}\left[-(\alpha + k) \pm \sqrt{(\alpha + k)^2 - \frac{4\alpha}{h}}\right]. \tag{4.7}$$

The initial conditions give

$$s + C_+ + C_- = 0, \tag{4.8a}$$

$$c + \lambda_+ C_+ + \lambda_- C_- = v_0, \tag{4.8b}$$

so that

$$C_\pm = \pm\frac{v_0 - c + s\lambda_\mp}{\lambda_+ - \lambda_-}. \tag{4.9}$$

From Eq. (4.4) I find

$$\ddot{x} = -a_{max} + \lambda_+^2 C_+ e^{\lambda_+ t} + \lambda_-^2 C_- e^{\lambda_- t}. \tag{4.10}$$

Then, substitution of Eqs. (4.7) and (4.9) in Eq. (4.10) gives

$$\ddot{x} = -a_{max}\left[1 - \frac{1}{2}\left(1 + \frac{\alpha-k}{\sqrt{(\alpha+k)^2 - \frac{4\alpha}{h}}}\right)e^{\lambda_+ t} - \frac{1}{2}\left(1 - \frac{\alpha-k}{\sqrt{(\alpha+k)^2 - \frac{4\alpha}{h}}}\right)e^{\lambda_- t}\right]. \qquad (4.11)$$

Eq. (4.11) can be rearranged as

$$\ddot{x} = -a_{max}\left[1 - \frac{1}{2}\left(e^{\lambda_+ t} + e^{\lambda_- t}\right) - \frac{1}{2}\left(\frac{\alpha-k}{\omega}\right)\left(e^{\lambda_+ t} - e^{\lambda_- t}\right)\right], \qquad (4.12)$$

where

$$\omega = \frac{1}{2}\sqrt{(\alpha+k)^2 - \frac{4\alpha}{h}}. \qquad (4.13)$$

If $(\alpha+k)^2 > \frac{4\alpha}{h}$ and thus $\omega$ is real, the acceleration decreases monotonically from 0 at $t = 0$ to $-a_{max}$ as $t \to \infty$ and consequently Eq. (4.3) remains valid. If $\omega$ is not real, the deceleration could exceed $a_{max}$. The stability condition $\alpha + 2k > \frac{2}{h}$ is generally not sufficient to make $\omega$ real. Note that when $k = \frac{1}{h}$ the stability condition is satisfied for any $\alpha > 0$ and $\omega$ is real; also the acceleration simplifies to $\ddot{x} = -a_{max}[1 - e^{-kt}]$.

The headway is

$$H(t) = x_l(t) - x_1(t) - D. \qquad (4.14)$$

After some manipulations, I find

$$H(t) = hv_0 - ha_{max}t + a_{max}\frac{h}{\alpha}[h(\alpha+k) - 1][1 - e^{-\gamma t}\cosh(\omega t)]$$

$$+ \frac{ha_{max}[3\alpha + k - h(\alpha+k)^2]}{2\alpha\omega} e^{-\gamma t}\sinh(\omega t), \qquad (4.15)$$

where

$$\gamma = \frac{\alpha+k}{2}. \qquad (4.16)$$

At $t = t_0 = \frac{v_0}{a_{max}}$ the leading vehicle velocity is zero. In Fig. 3, the headway (red curve) at this time is shown as a function of $\alpha$ for $k = k_{min} = max\left\{\left[\frac{1}{h} - \frac{\alpha}{2}\right], \left[2\sqrt{\frac{\alpha}{h}} - \alpha\right]\right\}$ (blue curve), which is the minimum value of $k$ to make $\omega$ real and the platoon string stable. Likewise, the velocity is

$$\dot{x}(t_0) = ha_{max}\left[1 - e^{-\gamma t_0}\left\{\cosh(\omega t_0) + \frac{[2\omega^2 + \alpha/h][2\gamma h - 1] - \alpha\gamma}{\alpha\omega}\sinh(\omega t_0)\right\}\right]. \tag{4.17}$$

Let

$$y(t) = x(t) - x(t_0) - H(t_0), \tag{4.18}$$

so that

$$H(t) = -y(t), \quad t > t_0. \tag{4.19}$$

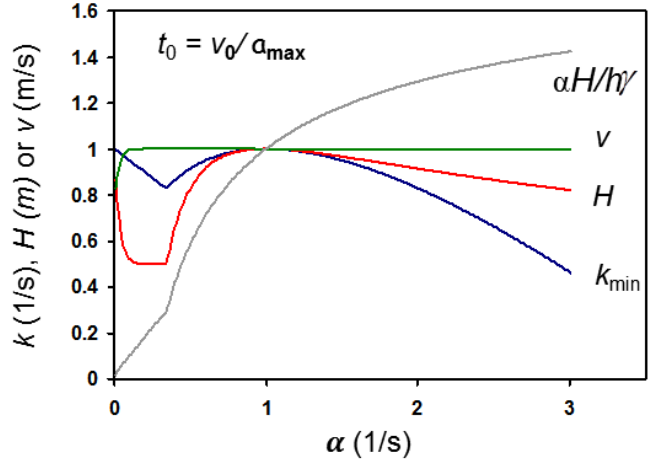

Fig. 3. The headway $H(t_0)$ of vehicle $n = 1$ (red curve) and velocity (green curve) at time $t_0 = \frac{v_0}{a_{max}}$ as a function of $\alpha$ with $k$ given by $k_{min}$ (blue curve), which is the larger of the string stability condition or the value of $k$ to make $\omega$ is real. $k_{min} = max\left\{\left[\frac{1}{h} - \frac{\alpha}{2}\right], \left[2\sqrt{\frac{\alpha}{h}} - \alpha\right]\right\}$. The leading vehicle decelerates to 0 at $a_{max} = 1$ m/s² from an initial velocity $v_0 = 32$ m/s. Vehicle 1 also starts at 32 m/s with a headway of 32 m. The gray curve is the maximum velocity vehicle 1 can have at $t_0$ and not have a final headway less than zero. For $t > t_0$ vehicle 1 approaches the stopped leading vehicle. The no-collisions condition is satisfied only for $\alpha > 1$ s⁻¹. The vehicle mechanical response is taken to be instantaneous.

The equation of motion for vehicle $n = 1$ as it approaches the stopped leading vehicle is

$$\ddot{y} + 2\gamma\dot{y} + \frac{\alpha}{h}y = 0, \tag{4.20}$$

with initial conditions

$$y(t_0) = -H(t_0), \tag{4.21a}$$

$$\dot{y}(t_0) = \dot{x}(t_0). \tag{4.21b}$$

If

$$\dot{y}(t_0) < \frac{\alpha|y(t_0)|}{h\gamma}, \tag{4.22}$$

then

$$y(t) < 0, \; t > t_0, \tag{4.23}$$

and no collision occurs ($H(t) > 0$). Eq. (4.23) can also be expressed as

$$v(t_0) < \frac{\alpha H(t_0)}{h\gamma}. \tag{4.24}$$

Note that value of $a_{max}$ cancels out except for the exponential terms which are negligible.

Simulation results are shown in Fig. 3 for large $v_0$ and $k = k_{min}$. When $\alpha > 1$ s$^{-1}$ and $h = 1$ s, the velocity (green curve) is less than $\frac{\alpha H(t_0)}{h\gamma}$ (gray curve), so that no collisions occur. The velocity is approximately 1 m/s for $a_{max} = 1$ m/s$^2$. If $k = \frac{1}{h}$, it straightforward to show that no collisions occur when $\alpha > \frac{1}{h}$ in agreement with the simulations. Further calculations for a platoon (Fig. 4) demonstrate that no collisions occur if $k > k_{min}$ and $\alpha > 1$ s$^{-1}$. Due to scaling, the more general statement is no collisions happen when

$$hk > 2\sqrt{h\alpha} - h\alpha, \; h\alpha > 1. \tag{4.25}$$

Subsequent vehicles in the platoon each follow a vehicle that remains with the limits and thus also remain within the limits.

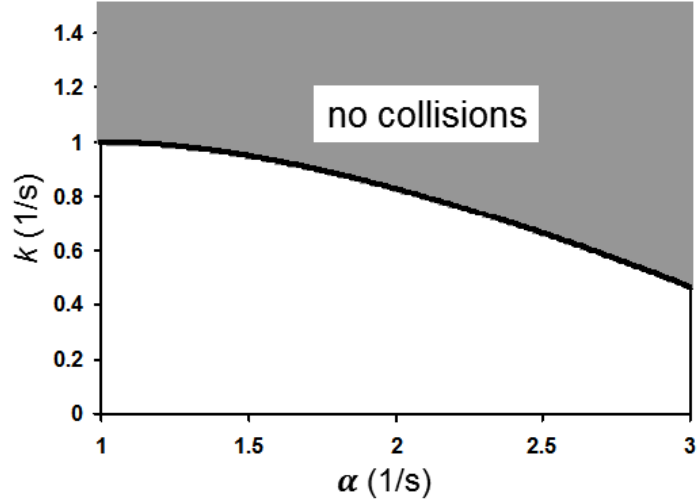

Fig. 4. The region in $\alpha - k$ space where the motion of the platoon following a leading vehicle that decelerates to zero velocity (Fig. 3) is stable and involves no collisions (gray). The lower boundary (black curve) is $k_{min}$ vs. $\alpha$. The vehicle mechanical response is taken to be instantaneous.

## 5. Simulations with mechanical response

In this section I consider a more general scenario where the mechanical response of each vehicle is not instantaneous. Instead it is described by a first-order time constant $\tau$ and an explicit delay $t_d$. Acceleration feedback is added to achieve string stability because the actual acceleration lags the desired acceleration [18]. The equation of motion is given by

$$\tau \frac{da_n}{dt} + a_n(t) = a_{dn}(t - t_d). \tag{5.1}$$

The desired acceleration is

$$a_{dn}(t) = min\{a_{max}, max[-a_{max}, A_n(t) - \xi a_n(t)]\}, \qquad (5.2)$$

where $A_n(t)$ is given by Eq. (2.2). The acceleration feedback gain is denoted by $\xi$.

Now I examine simulations where the leading vehicle decelerates at $a_{max}$ from $v_{max}$ to 0 and remains stopped. In Fig. 5, I show the region of the $\tau - \xi$ plane (for fixed $t_d$) where no collisions occur. The lower boundary (gray region) is determined by the string-stable condition. The darker region corresponds to the "stable but with unacceptable acceleration oscillations" [22] condition. Note that $\xi$ = 0.75, a typical gain for $\tau = t_d = 0.3$ s, is too large for small $\tau$. Similarly, no single value of $\xi$ guarantees no collisions for all $\tau$ when $t_d = 0$ as shown in Fig. 6. In [22] it was shown the gain $\xi$ should be chosen near, but not in, the darker region to have the quickest reduction of a perturbation as $n \to \infty$. These results indicate that, to obtain optimal performance, the gain $\xi$ might have to be dynamically varied as the mechanical response changes with engine speed.

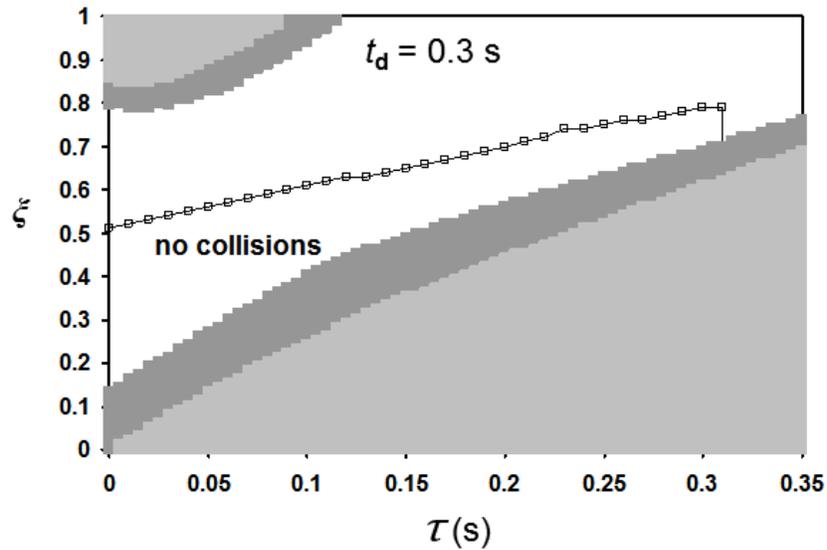

Fig. 5. Acceleration feedback gain $\xi$ vs. $\tau$ for $t_d$ = 0.3 s. The leading vehicle decelerates from 32 m/s to 0 at $a_{max}$ = 1 ms$^{-2}$ and remains stopped. The light gray regions are unstable. The dark gray regions are string stable but have unacceptable oscillations in acceleration. The squares are the smallest $\xi$ for which a collision occurs in the stable region. Only the region below the squares and above the gray region is stable and without collisions. $\alpha = 2$ s$^{-1}$, $k = 1$ s$^{-1}$.

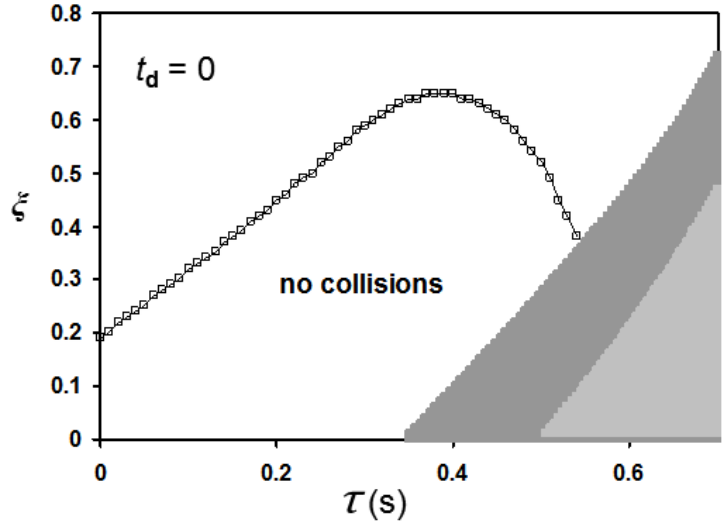

Fig. 6. Acceleration feedback gain $\xi$ vs. $\tau$ for $t_d$ = 0. The leading vehicle decelerates from 32 m/s to 0 at $a_{max}$ = 1 ms$^{-2}$ and remains stopped. The light gray regions are unstable. The dark gray regions are string stable but have unacceptable oscillations in acceleration. The squares are the minimum $\xi$ for which a collision occurs in the stable region. Only the region below the squares and above the gray region is stable and without collisions. $\alpha$ = 2 s$^{-1}$, $k$ = 1 s$^{-1}$.

Fig. 7 depicts, for several values of the acceleration feedback gain, the regions in the $\tau - t_d$ plane in which collisions can occur. The string-stable, collision-free region for $\xi = 0.75$ is small and is enclosed by a closed curve that only partially overlaps the collision-free regions for smaller gains (Figs. 7a and 7b). There is no overlap with the string-stable region, which is also free of collisions, for $\xi = 0$ (Fig. 7c). If the leading vehicle has the velocity profile of Fig. 1, which is less demanding, the string-stable, collision-free region for $\xi = 0.75$ is larger as shown in Fig. 8a. Even so, the overlaps with the corresponding regions for lesser gains is small or zero (Figs. 8b, 8c and 8d). Both of these figures indicates the potential need for adjusting the gain as engine speed changes and the mechanical response varies.

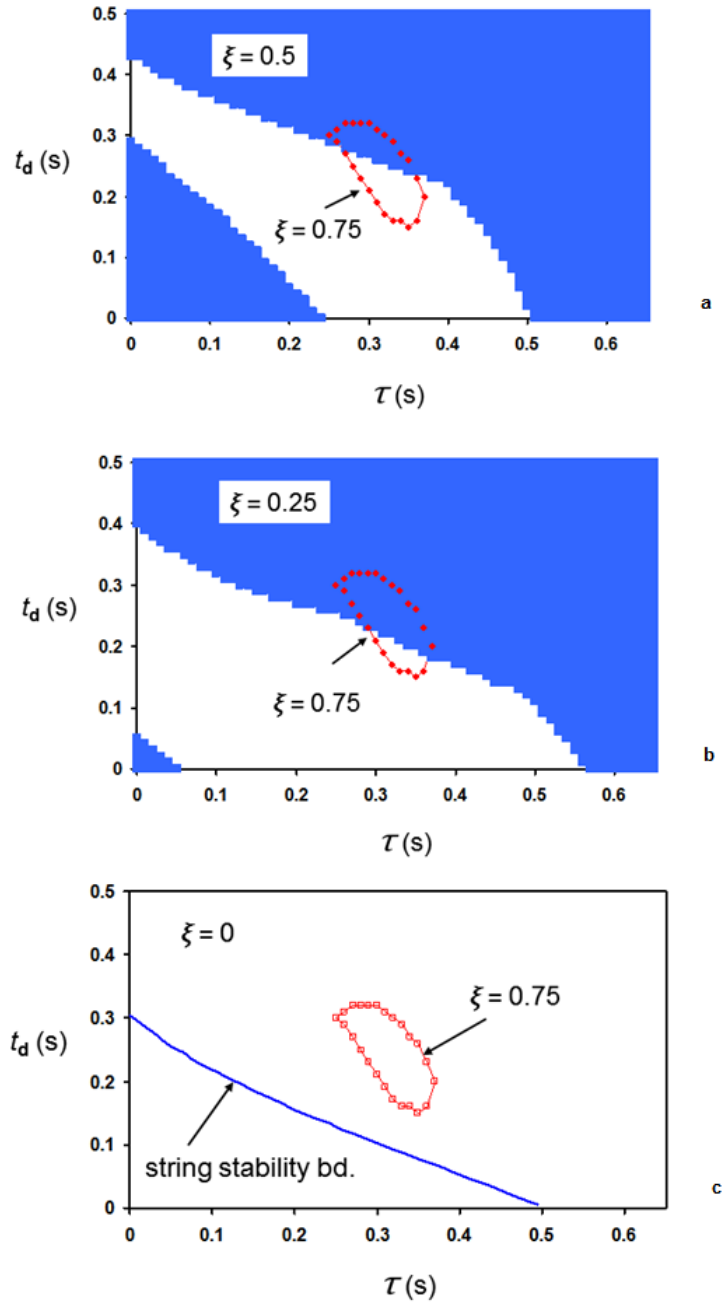

Fig. 7. Regions in the $\tau - t_d$ plane in which collisions can occur (blue) for simulations where the leading vehicle decelerates from 32 m/s to 0 at 1 ms$^{-2}$ and remains stopped. The red curve encloses the collision-free, string-stable region for $\xi = 0.75$. (a) The plot for $\xi = 0.5$; (b) the plot for $\xi = 0.25$;. (c) the comparison to the string stability boundary for $\xi = 0$. $\alpha = 2$/s, $k$ = 1/s and $h$ = 1 s.

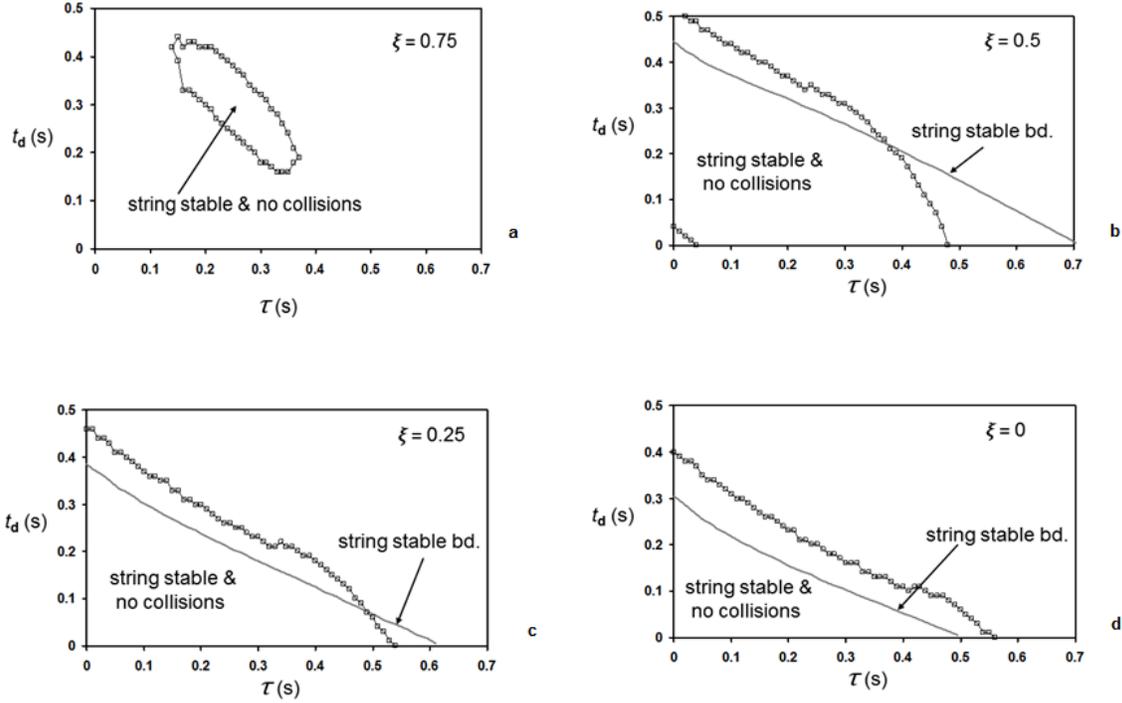

Fig. 8. Regions in the $\tau - t_d$ plane that are string stable and collision free for (a) $\xi = 0.75$; (b) $\xi = 0.5$; (c) $\xi = 0.25$; (d) $\xi = 0$. The gray curves are the string stability boundaries and the connected squares are the boundaries for the regions where collisions can occur. The simulations were done for the leading vehicle speed profile of Fig. 1. $\alpha = 2$/s, k = 1/s and h = 1 s.

### 6. Stability against perturbations

In this section I study the stability of a platoon when a vehicle is perturbed from its equilibrium velocity by a finite amount, *i. e.*, large enough to produce acceleration (or deceleration) of magnitude $a_{max}$. The leading vehicle moves at a steady velocity $v_0$. All vehicles in the platoon following the leading vehicle have headway $hv_0$ and vehicles *n* =2, 3… begin with velocity $v_0$. Vehicle *n* =1 is perturbed to a new velocity $v_1$, which is chosen to be almost, but not quite, extreme enough to cause a collision. The motion of the vehicles in the platoon is calculated to demonstrate stability.

For these calculations, the n = 1 vehicle initially has velocity $v_1$ = 24 m/s and all other vehicles begin at $v_0$ = 20 m/s. The parameters of the control algorithm are $\alpha = 2$/s, k = 1/s and h = 1 s with $a_{max} = 1$ m/s$^2$ and $\tau = t_d = 0.3$ s with $\xi = 0.75$. In Fig. 9 the velocity and acceleration of n = 1, 2 and 3 are shown (panels a, b, c respectively). Because vehicles 1 and 2 do not begin in the equilibrium state, both are constrained by $a_{max}$. [Vehicle 2 is not in equilibrium because the acceleration (for instantaneous response) at t = 0 is not zero.] Subsequent vehicles n = 3, 4… are initially at equilibrium and thus only approach, but do not reach, the limits.

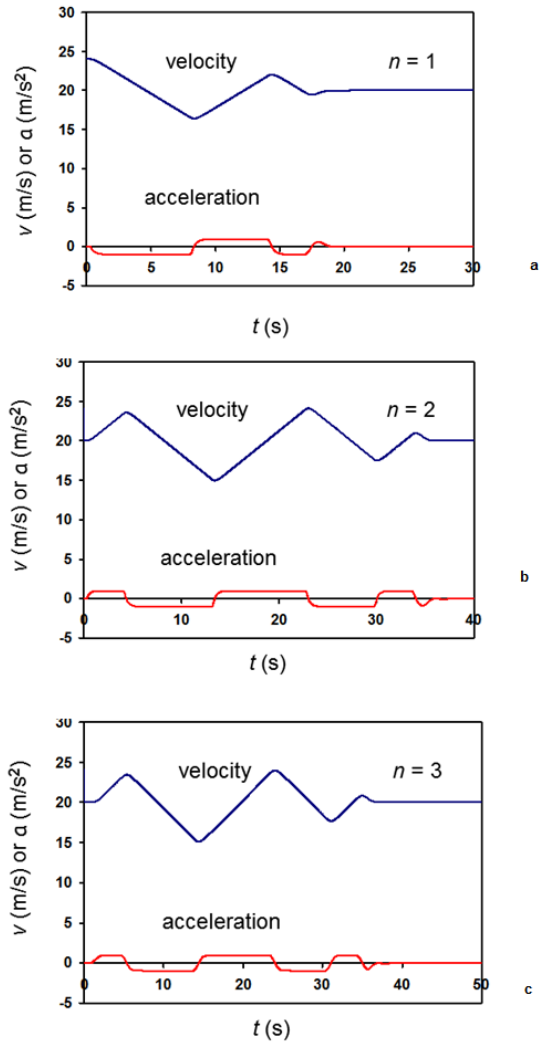

Fig. 9. Velocity (blue) and acceleration (red) as functions of time for vehicles (a) $n$ =1; (b) $n$ =2; (c) $n$ =3. The leading vehicle moves at a steady 20 m/s. Initially all vehicles in the platoon have a headway of 20 m and a velocity of 20 m/s, except vehicle 1 which has an initial velocity of 24 m/s. $\tau = t_d = 0.3$ s and $\xi = 0.75$. $\alpha = 2$/s, k = 1/s and h = 1 s.

Furthermore the size of the perturbation decreases as *n* becomes larger, as can be seen in Fig. 10 where the velocities for *n* = 1, 5, 10 and 25 are shown. Similar results are found for $v_1 < v_0$.

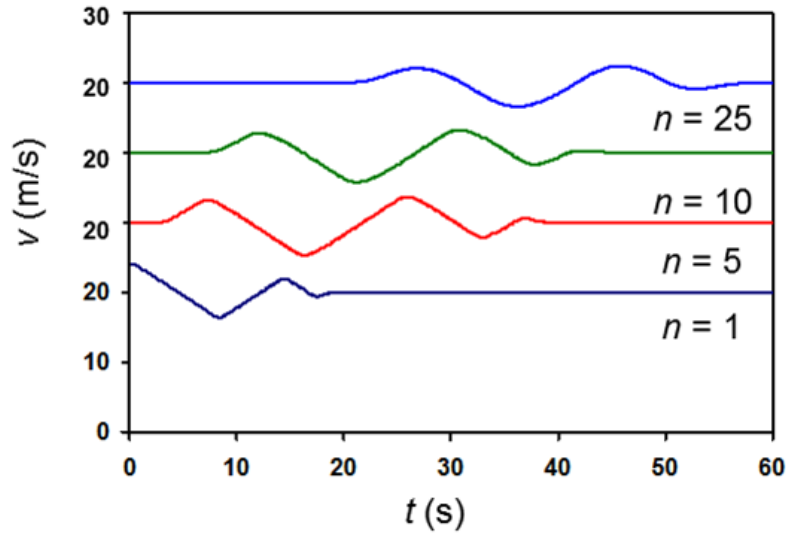

Fig. 10. Velocity *vs.* time for vehicles $n$ =1, 5, 10 and 25 offset by 10 m/s each. The initial conditions are the same as in Fig. 9. $\tau = t_d = 0.3$ s and $\xi = 0.75$. $\alpha = 2$/s, *k* = 1/s and *h* = 1 s.

Results from simulations done for perturbations of both position $\Delta x$ and velocity $\Delta v$ of vehicle 1 from its initial equilibrium state are presented in Fig. 11. The region of stable motion with no collisions is shown in the $\Delta x - \Delta v$ plane for $\tau = t_d = 0.3$ s and $\xi = 0.75$ (enclosed by the red curves) and for instantaneous response (enclosed by the blue curves). The latter region is somewhat larger as expected. In the initial state all vehicles have velocity 20 m/s and are in equilibrium positions (except vehicle 1) with the leading vehicle moving at 20 m/s. Fig. 11 implies that a platoon can recover stable motion from non-equilibrium initial states that induce maximum acceleration or deceleration.

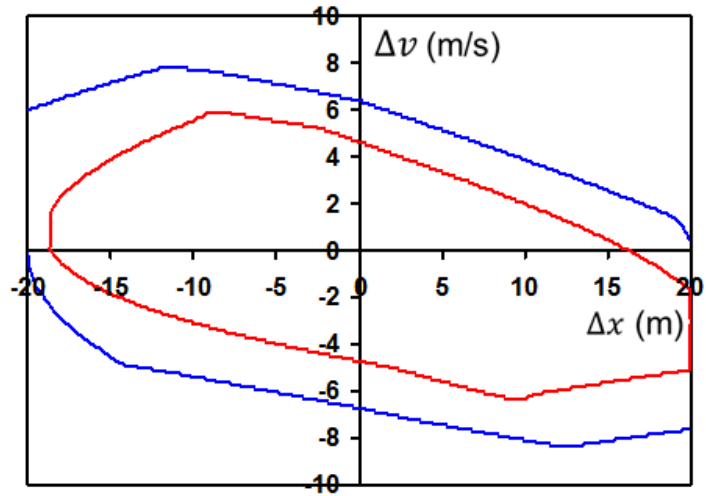

Fig. 11. The boundaries of stable motion with no collisions where the $n=1$ vehicle is perturbed from its initial equilibrium state by $\Delta x$ in position and $\Delta v$ in velocity. All other vehicles in the platoon have velocity 20 m/s and headway 20 m initially. The leading vehicle moves at 20 m/s. The blue curve is for instantaneous response and no acceleration feedback. The red curve is for $\tau = t_d = 0.3$ s and $\xi = 0.75$. $\alpha = 2$/s, $k$ = 1/s and $h$ = 1 s.

## 7. Conclusions

In this paper I have examined the effects of imposing maximum acceleration and deceleration limits on an otherwise string-stable platoon of ACC vehicles. The limits are realistic values of comfortable acceleration and decelerations due strictly to the powertrain. Braking is not included in the analysis. Application of the brakes is reserved for emergency situations only.

Throughout the paper, the assumptions made are (1) the platoon is string-stable for small perturbations (in regions of parameter space labeled "stable") and (2) the initial state of the platoon is the equilibrium state (except in Sec. 6). The requirement for a suitable range of control law parameters is that no collisions occur, $i.e.$, the headway for any vehicle never becomes zero or negative.

In Sec. 2, I showed that for a simple model with relative velocity gain $k = \frac{1}{h}$ and instantaneous mechanical response, if the acceleration of vehicle $n$-1 remains in the interval $[-a_{max}, a_{max}]$, then the acceleration of vehicle $n$ also remains in this interval, independent of the sensitivity $\alpha$. This implies that if the vehicle leading a platoon does not exceed the limits, no vehicle in the platoon does.

Simulations for the simple model where the leading vehicle alternatively accelerates and decelerates at $a_{max}$ show that if $k \neq \frac{1}{h}$, there is a region in the $\alpha - k$ plane that even string-stable platoons experience collisions. The condition that no collisions happen when the vehicle in front decelerates at $a_{max}$ to zero velocity and remains stopped has been determined to be $\alpha > \frac{1}{h}$ and $k > 2\sqrt{\frac{\alpha}{h}} - \alpha$.

If the model is generalized to include mechanical response, simulations demonstrate that for $\alpha = \frac{2}{h}$ and $k = \frac{1}{h}$ there exist acceleration feedback gains that are in the stable zone with no oscillations [22] such that the platoon remains stable when the limits on acceleration are imposed. Simulations with different mechanical responses (such as caused by changes in engine speed) demonstrate that the gain might have to be dynamically varied to obtain optimal performance.

Finally, simulations show that stability can be found when the first vehicle's velocity or position (or both) is perturbed from its equilibrium value by a large enough amount to induce acceleration and deceleration at $a_{max}$ without incurring collisions.